\documentclass{article}
\usepackage{amsmath,amsthm}
\usepackage{amssymb,latexsym}
\usepackage[mathscr]{eucal}
\usepackage{setspace}
\usepackage{graphics}
\usepackage{array}

\newcommand{\al}{\alpha}
\newcommand{\na}{\nabla}
\newcommand{\pa}{\partial}

\newcommand{\eps}{\epsilon}
\newcommand{\lp}{\left(}
\newcommand{\rp}{\right)}
\newcommand{\lb}{\left[}
\newcommand{\rb}{\right]}

\newcommand{\la}{\langle}
\newcommand{\ra}{\rangle}
\newcommand{\bs}{\boldsymbol}
\newcommand{\ti}{\times}
\newcommand{\be}{\begin{equation}}
\newcommand{\ee}{\end{equation}}

\newcommand{\ihat}{\bf\hat{i}}

\newcommand{\C}{\mathcal{C}}

\begin{document}

\begin{center}
\huge{\bf Plasma Relaxation and Topological Aspects in Hall Magnetohydrodynamics}

\vspace{.3in}

\large{B.K. Shivamoggi\\
University of Central Florida\\
Orlando, FL 32816-1364}
\end{center}

\vspace{.3in}

\noindent\Large{\bf Abstract}\\

\large Parker's formulation of isotopological plasma relaxation process in magnetohydrodynamics (MHD) is extended to Hall MHD. The torsion coefficient $\al$ in the Hall MHD Beltrami condition turns out now to be proportional to the ``{\it potential vorticity}." The Hall MHD Beltrami condition becomes equivalent to the ``{\it potential vorticity}" conservation equation in two-dimensional (2D) hydrodynamics if the Hall MHD Lagrange multiplier $\beta$ is taken to be proportional to the ``{\it potential vorticity}" as well. The winding pattern of the magnetic field lines in Hall MHD then appears to evolve in the same way as ``{\it potential vorticity}" lines in 2D hydrodynamics.

\pagebreak

\noindent\Large{\bf 1. Introduction}\\

\large A significant class of exact solutions of the equations governing magnetohydrodynamics (MHD) emerges under the Beltrami condition - the local current density is proportional to the magnetic field - the {\it force-free} state (Lundquist \cite{Lun}, Lust and Schluter \cite{Lus}). These Beltrami solutions turned out to correlate well with real plasma behavior (Priest and Forbes \cite{Pri}, Schindler \cite{Sch}). Parker \cite{Par} - \cite{Par3} showed that, in certain plasma relaxation processes, the Beltrami condition is indeed equivalent to the vorticity conservation equation in two-dimensional (2D) hydrodynamics (and the Lagrange multiplier $\al$ turned out to be proportional to vorticity).

In a high-$\beta$ plasma, on length scales in the range $d_e < \ell < d_i$, where $d_s$ is the skin depth, $d_s \equiv c/\omega_{p_s}$, $s = i, e$ ($i$ and $e$ referring to the ions and electrons, respectively), the electrons decouple from the ions. This results in an additional transport mechanism for the magnetic field via the Hall current (Sonnerup \cite{Son}), which is the ion-inertia contribution in Ohm's law. The Hall effect leads to the generation of whistler waves whose,
\begin{itemize}
  \item frequency lies between ion-cyclotron and electron-cyclotron frequencies $\omega_{c_i}$ and $\omega_{c_e}$, respectively,
  \item phase velocity exceeds that of Alfv\'{e}n waves for wavelengths parallel to the applied magnetic fields less than $d_i$.
\end{itemize}

Further, the decoupling of ions and electrons in a narrow region around the magnetic neutral point (where the ions become unmagnetized while the electrons remain magnetized) allows for rapid electron flows in the ion-dissipation region and hence a faster magnetic reconnection process in the Hall MHD regime (Mandt et al. \cite{Man}).

The purpose of this paper is to extend Parker's \cite{Par} - \cite{Par3} considerations to Hall MHD and investigate the evolution of the winding pattern of the magnetic field lines in Hall MHD.

\vspace{.3in}

\noindent\Large{\bf 2. Beltrami States in Hall MHD}\\

\large The Hall MHD equations (which were formulated by Lighthill \cite{Lig} following his far-sighted recognition of the importance of the Hall term in the generalized Ohm's law) are (in usual notations),
\be\tag{1}
\frac{\pa \bs\Omega_i}{\pa t} = \na \ti \lp {\bf v}_i \ti \bs\Omega_i \rp
\ee
\be\tag{2}
\frac{\pa {\bf A}}{\pa t} = \frac{1}{c} {\bf v}_i \ti {\bf B} - \frac{1}{nec} ~{\bf J} \ti {\bf B}
\ee
where $n$ is the number density of ions (or electrons) and $\bs\Omega_i$ is the generalized vorticity,
\be\tag{3}
\bs\Omega_i \equiv \bs\omega_i + \bs\omega_{c_i}, ~\bs\omega_i \equiv \na \ti {\bf v}_i, ~\bs\omega_{c_i} \equiv \frac{e {\bf B}}{m_i c}.
\ee
Here, we have considered an incompressible, two-fluid, quasi-neutral plasma and have neglected the electron inertia.

Equations (1) and (2) have the Hamiltonian formulation (Shivamoggi \cite{Shi}),
\be\tag{4}
H = \frac{1}{2} \int\limits_V \lb \bs\psi_i \cdot \bs\Omega_i + \frac{1}{c} {\bf A} \cdot \lp {\bf J} - ne {\bf v}_i \rp \rb d V
\ee
where,
\be\tag{5}
m_i n {\bf v}_i \equiv \na \ti \bs\psi_i
\ee
and $V$ is the volume occupied by the plasma.\footnote{(5) implies
\be\notag
\frac{\pa n}{\pa t} = 0
\ee
in accord with the assumption that the plasma is incompressible.} Further, we have put $|\bs\psi_i| = 0$ on the boundary $\pa V$, and have rendered $\bs\psi_i$ unique by imposing the gauge condition
\be\tag{6}
\na \cdot \bs\psi_i = 0.
\ee

We choose $\lp \bs\Omega_i, {\bf A} \rp$ to be the canonical variables, and take
\be\tag{7}
J \equiv \lp
\begin{matrix}
- \na \ti \lp \displaystyle\frac{\bs\Omega_i}{m_i n} \ti \lp \na \ti \lp \cdot \rp \rp \rp & {\bf 0}\\
{\bf 0} & \displaystyle\frac{c {\bf B}}{n e} \ti \lp \cdot \rp
\end{matrix}
\rp
\ee
as a $\lp \bs\Omega_i, {\bf A} \rp$-dependent differential operator which produces a skew-symmetric transformation of vector functions vanishing on $\pa V$ and satisfies a closure condition on an associated symplectic two-form (Olver \cite{Olv}).

The Hamilton equations are then
\be\tag{8}
\lp
\begin{matrix}
\displaystyle\frac{\pa \bs\Omega_i}{\pa t}\\
\\
\displaystyle\frac{\pa {\bf A}}{\pa t}
\end{matrix}
\rp = J \lp
\begin{matrix}
\displaystyle\frac{\delta H}{\delta \bs\Omega_i}\\
\\
\displaystyle\frac{\delta H}{\delta {\bf A}}
\end{matrix}
\rp
\ee
which are just equations (1) and (2). Here, $\delta H/\delta {\bf q}$ is the variational derivative.

The Casimir invariants for Hall MHD are solutions of the equations,
\be\tag{9}
J \lp
\begin{matrix}
\displaystyle\frac{\delta \C}{\delta \bs\Omega_i}\\
\\
\displaystyle\frac{\delta \C}{\delta {\bf A}}
\end{matrix}
\rp = \lp
\begin{matrix}
{\bf 0}\\
\\
{\bf 0}
\end{matrix}
\rp.
\ee

It may be verified that two such solutions are
\be\tag{10}
\lp
\begin{matrix}
\displaystyle\frac{\delta \C_{(1)}}{\delta \bs\Omega_i}\\
\\
\displaystyle\frac{\delta \C_{(1)}}{\delta {\bf A}}
\end{matrix}
\rp = \lp
\begin{matrix}
{\bf 0}\\
\\
{\bf B}
\end{matrix}
\rp
\ee
or
\be\tag{11}
\C_{(1)} = \int\limits_V {\bf A} \cdot {\bf B} ~d V
\ee
as with classical MHD, and
\be\tag{12}
\lp
\begin{matrix}
\displaystyle\frac{\delta \C_{(2)}}{\delta \bs\Omega_i}\\
\\
\displaystyle\frac{\delta \C_{(2)}}{\delta {\bf A}}
\end{matrix}
\rp = \lp
\begin{matrix}
\displaystyle\frac{e {\bf A}}{m_i c} + {\bf v}_i\\
\\
\displaystyle\lp\frac{e}{m_i c} \rp^2 {\bf B}
\end{matrix}
\rp
\ee
or
\be\tag{13}
\C_{(2)} = \int\limits_V \lp \frac{e {\bf A}}{m_i c} + {\bf v}_i \rp \cdot \bs\Omega_i ~d V.
\ee
$\C_{(1)}$ is the total magnetic helicity and $\C_{(2)}$ is the total generalized ion cross helicity.

A significant class of exact solutions of the Hall MHD equations (1) and (2) emerges as the end result of the isotopological energy-lowering Beltramization process. Thus, minimization of $H$, keeping $\C_{(1)}$ fixed, gives
\be\tag{14}
\frac{\delta H}{\delta {\bf A}} = \lambda_{(1)} \frac{\delta \C_{(1)}}{\delta {\bf A}}
\ee
or
\be\tag{15}
\frac{1}{c} \lp {\bf J} - ne {\bf v}_i \rp = \lambda_{(1)} {\bf B}
\ee
which is the pseudo-force-free state.

On the other hand, minimization of $H$, keeping $\C_{(2)}$ fixed, gives
\be\tag{16}
\frac{\delta H}{\delta \bs\Omega_i} = \lambda_{(2)} \frac{\delta \C_{(2)}}{\delta \bs\Omega_i}
\ee
or
\be\tag{17}
m_i n {\bf v}_i = \lambda_{(2)} \bs\Omega_i
\ee
which is the generalized Alfv\'{e}nic state.

Combining (15) and (17), we obtain for the Hall MHD Betrami state (Turner \cite{Tur}),
\be\tag{18}
\frac{m_i}{e} \na \ti {\bf B} - \lp \lambda_{(1)} \frac{m_i}{e} + \frac{e}{m_i c} \lambda_{(2)} \rp {\bf B} = \lambda_{(2)} \bs\omega_i.
\ee

\vspace{.3in}

\noindent\Large{\bf 3. Plasma Relaxation in an Applied Uniform Magnetic Field}\\

\large Consider now, following Parker \cite{Par} - \cite{Par3}, a plasma in an applied uniform magnetic field ${\bf B}_0 = B_0 {\ihat}_z$ and confined between two infinite parallel planes $z = 0 ~\text{and} ~L$, which relaxes\footnote{In this process, the magnetic field lines extending between the planes $z = 0 ~\text{and} ~L$ are wrapped around and intermixed by the motion of their foot points on these planes (Parker \cite{Par} - \cite{Par3}).} isotopologically toward the lowest available energy state described by equation (18) written in the form
\be\tag{19}
\na \ti {\bf B} = \al {\bf B} + \beta \bs\omega_i.
\ee
The MHD Lagrange multiplier $\al$ may be interpreted as the torsion coefficient while $\beta$ is the Hall MHD Lagrange multiplier.

Suppose this process exhibits slow variations in the z-direction, characterized by the slow spatial scale,
\be\tag{20}
\xi \equiv \eps z, ~\eps \ll 1.
\ee
Let the magnetic field involved in this process be given by
\be\tag{21}
{\bf B} = \la \eps B_0 b_x, ~\eps B_0 b_y, ~B_0 \lp 1 + \eps b_z \rp \ra
\ee
and the Lagrange multipliers $\al$ and $\beta$ be given by
\be\tag{22}
\al = \eps a, ~\beta = \eps b.
\ee

Using (20) - (22), equation (17) may be written as
\be\tag{23a}
v_x = \sigma \lp c_1 \eps b_x + \omega_x \rp
\ee
\be\tag{23b}
v_y = \sigma \lp c_1 \eps b_y + \omega_y \rp
\ee
\be\tag{23c}
v_z = \sigma \lb c_1 \lp 1 + \eps b_z \rp + \eps \omega_z \rb.
\ee
The out-of-plane (or {\it toroidal}) ion flow $\lp v_z \not= 0 \rp$ is peculiar to Hall MHD. Here, $\sigma$ and $c_1$ are appropriate constants. Equation (19) leads to
\be\tag{24a}
\frac{\pa b_z}{\pa y} - \eps \frac{\pa b_y}{\pa \xi} = \eps a b_x + \eps b \omega_x
\ee
\be\tag{24b}
\eps \frac{\pa b_x}{\pa \xi} - \frac{\pa b_z}{\pa x} = \eps a b_y + \eps b \omega_y
\ee
\be\tag{24c}
\frac{\pa b_y}{\pa x} - \frac{\pa b_x}{\pa y} = a \lp 1 + \eps b_z \rp + \eps b \omega_z
\ee
and the divergence-free condition on {\bf B} leads to
\be\tag{25}
\frac{\pa b_x}{\pa x} + \frac{\pa b_y}{\pa y} + \eps \frac{\pa b_z}{\pa \xi} = 0.
\ee

On the other hand, taking the divergence of equation (19), we obtain
\be\tag{26}
{\bf B} \cdot \na \al + \bs\omega_i \cdot \na \beta = 0
\ee
which, on using (20) - (22), leads to
\be\tag{27}
b_x \frac{\pa a}{\pa x} + b_y \frac{\pa a}{\pa y} + \lp 1 + \eps b_z \rp \frac{\pa a}{\pa \xi} + \omega_x \frac{\pa b}{\pa x} + \omega_y \frac{\pa b}{\pa y} + \eps \omega_z \frac{\pa b}{\pa \xi} = 0.
\ee

Equations (24a) and (24b) imply,
\be\tag{28}
b_z \sim O \lp \eps \rp.
\ee
Using (28), equation (25) leads to, to $O \lp 1 \rp$,
\be\tag{29}
b_x = \frac{\pa \psi}{\pa y}, ~b_y = -\frac{\pa \psi}{\pa x}
\ee
for some magnetic flux function $\psi = \psi \lp x, y \rp$.

Using (29), we obtain from (23), to $O \lp \eps \rp$,
\be\tag{30a}
\frac{\pa v_x}{\pa y} = \sigma \lp c_1 \eps \frac{\pa^2 \psi}{\pa y^2} + \eps \frac{\pa^2 v_z}{\pa y^2} \rp
\ee
\be\tag{30b}
\frac{\pa v_y}{\pa x} = \sigma \lp -c_1 \eps \frac{\pa^2 \psi}{\pa x^2} - \eps \frac{\pa^2 v_z}{\pa x^2} \rp
\ee
and hence,
\be\tag{31}
\omega_z \equiv \frac{\pa v_y}{\pa x} - \frac{\pa v_x}{\pa y} = -\sigma \eps \lp c_1 \na^2 \psi + \na^2 v_z \rp
\ee
where,
\be\notag
\na^2 \equiv \frac{\pa^2}{\pa x^2} + \frac{\pa^2}{\pa y^2}.
\ee

Next, using (29), equation (24c) leads to, to $O \lp 1 \rp$,
\be\tag{32}
a = -\na^2 \psi.
\ee
Using (31), and putting
\be\tag{33}
\omega_z \equiv \eps \sigma c_1 \omega
\ee
equation (32) leads to
\be\tag{34}
a = q \equiv \omega + \frac{1}{c_1} \na^2 v_z
\ee
implying that the torsion coefficient $\al$ is proportional to the ``{\it potential vorticity}" $q$ in Hall MHD.

On the other hand, using (23) and (34), equation (27) leads to, to $O \lp \eps \rp$,
\be\tag{35}
\eps \sigma c_1 \frac{\pa q}{\pa \xi} + v_x \frac{\pa q}{\pa x} + v_y \frac{\pa q}{\pa y} + \sigma \eps \lb \lp q - c_1 b \rp, v_z \rb = 0
\ee
where,
\be\notag
\lb f, g \rb \equiv \frac{\pa f}{\pa x} \frac{\pa g}{\pa y} - \frac{\pa f}{\pa y} \frac{\pa g}{\pa x}.
\ee
If we take the Hall MHD Lagrange multiplier $b$ also to be proportional to the ``{\it potential vorticity}" $q$, i.e.,
\be\tag{36}
b = \frac{1}{c_1} q
\ee
equation (35) becomes the ``{\it potential vorticity}" conservation equation in 2D hydrodynamics (on identifying $\xi$ with $t$),
\be\tag{37}
\eps \sigma c_1 \frac{\pa q}{\pa \xi} + v_x \frac{\pa q}{\pa x} + v_y \frac{\pa q}{\pa y} = 0.
\ee
Thus, the Beltrami condition (19) in Hall MHD becomes equivalent to the ``{\it potential vorticity}" conservation equation in 2D hydrodynamics if the Hall MHD Lagrange multiplier $\beta$ is taken to be proportional to the ``{\it potential vorticity}" $q$ as well.\footnote{(36) is {\it sufficient} but not necessary to obtain equation (37).} (34) then implies that the winding pattern of the magnetic field lines in Hall MHD evolves in the same way as ``{\it potential vorticity}" lines in 2D hydrodynamics.

\vspace{.3in}

\noindent\Large{\bf 4. Discussion}\\

\large In this paper, we have extended Parker's \cite{Par} - \cite{Par3} formulation of isotopological plasma relaxation process in MHD to Hall MHD. The torsion coefficient $\al$ in the Hall MHD Beltrami condition turns out now to be proportional to the ``{\it potential vorticity}." The Hall MHD Beltrami condition becomes equivalent to the ``{\it potential vorticity}" conservation equation in 2D hydrodynamics if the Hall MHD Lagrange multiplier $\beta$ is taken to be proportional to the ``{\it potential vorticity}" as well. The winding pattern of the magnetic field lines in Hall MHD then appears to evolve in the same way as ``{\it potential vorticity}" lines in 2D hydrodynamics. The analogy between a smooth, continuous magnetic field in Hall MHD and 2D hydrodynamics as in ordinary MHD (Parker \cite{Par3}) implies that the current sheets seem to have the same role in the development of Hall MHD equilibria as they do in the MHD case.
 
\vspace{.3in}

\noindent\Large{\bf Acknowledgments}\\

\large This work was a result of my participation at the {\it International Astrophysics Forum, Alpbach,} 2011. I am thankful to Professor Eugene Parker for helpful suggestions and giving me access to ref. \cite{Par3} prior to publication and Professors Manfred Leubner and Zoltan Voros for their hospitality.

\end{document}